\documentclass{ws-procs975x65}

\begin{document}



\title{TWO TEMPERATURE ACCRETION FLOWS AROUND ROTATING BLACK HOLES AND DETERMINING THE KERR PARAMETER OF SOURCES
}

\author{BANIBRATA MUKHOPADHYAY
}

\address{Department of Physics,
Indian Institute of Science, 
Bangalore-560012, India 
\email{bm@physics.iisc.ernet.in}}

%


\def\lsim{\lower.5ex\hbox{$\; \buildrel < \over \sim \;$}}
\def\gsim{\lower.5ex\hbox{$\; \buildrel > \over \sim \;$}}

\def\ch{\lower-0.55ex\hbox{--}\kern-0.55em{\lower0.15ex\hbox{$h$}}}
\def\lh{\lower-0.55ex\hbox{--}\kern-0.55em{\lower0.15ex\hbox{$\lambda$}}}

\begin{abstract}
We model two temperature viscous accretion flows in the sub-Keplerian,
optically thin, regime around rotating black holes including important 
radiation effects self-consistently. The model successfully explains 
observed luminosities from ultra-luminous to under-luminous sources and 
predicts the spin parameter of black holes.
\end{abstract}

\bodymatter

\section{Introduction}\label{intro}

\enlargethispage*{6pt}

It is well known that the low-hard state of Cyg~X-1 can not be explained\cite{ls75}
by the Keplerian accretion disk\cite{ss73}.
Therefore, Eardley, Lightman \& Shapiro\cite{els75} initiated to model 
the two temperature hot accretion flow. Later,
Muchotrzeb \& Paczy\'nski\cite{mp82} introduced the idea of sub-Keplerian,
transonic accretion,
which was later improved by other authors\cite{a88,c89,ny95,m03}, by discussing
importance of advection. 
Most of them introduced various cooling mechanisms, e.g., blackbody,
bremsstrahlung, synchrotron, inverse-Compton radiation, appropriately
according to their models. However, none of them  attempted to 
understand the variation of cooling/advective efficiency in the flow while infalling 
towards the black hole. Generally, it is expected that far away from the black hole 
the flow to be relatively cooler, while in the vicinity of the black hole it is hotter.

We, in the approximation of optically thin two temperature flow, plan to understand how the
flow behavior, in the light of advective efficiency, changes while infalling
towards a rotating black hole. This successfully explains luminosities of 
observed low to high luminous sources, in the framework of a single model,
which has not been attempted yet. In reproducing the correct luminosity of 
a source, the present model also predicts the spin parameter of the black hole at the
center.

\section{Model}\label{model}

The optically thin flow is assumed geometrically not to be thick enough 
so that the disk could be vertically averaged.
All the variables used here have their usual meanings and are expressed throughout 
in conventional dimensionless units, unless stated otherwise (see Rajesh \& 
Mukhopadhyay\cite{rm10}
for details). The equations of mass and momentum conservation are same as of previous
work\cite{mg03}. The proton and electron energy equations are given below as
\begin{eqnarray}
\nonumber
\frac{\vartheta h(x)}{\Gamma_{3} - 1} \left(\frac{dP}{dx} - \Gamma_{1} \frac{P}{\rho} \frac{d \rho}{dx}\right)= \,Q^{+} - Q_{ie},\,\,
\frac{\vartheta h(x)}{\Gamma_{3} - 1} \left(\frac{dP_{e}}{dx} - \Gamma_{1} \frac{P_{e}}{\rho} \frac{d \rho}{dx}\right) = Q_{ie} -Q^{-},\\
\label{eni}
\end{eqnarray}
when the Coulomb coupling $Q_{ie}$ is
given by
\begin{eqnarray}
\nonumber
q_{ie} = Q_{ie}\frac{c^{11}}{hG^4 M^3}=\frac{8 (2 \pi)^{1/2}e^4 n_i n_e}{m_i m_e}\left(\frac{T_e}{m_e} + \frac{T_i}{m_i}\right)^{-3/2} \ln (\Lambda) \left(T_i  -T_e \right){\rm erg/cm^3/sec},\\
\label{qie}
\end{eqnarray}
where $n_i$ and $n_e$ respectively denote number densities of ion and electron,
$e$ the electron charge,
ln($\Lambda$) the Coulomb logarithm, and
total heat radiated away ($Q^-$) by the
bremsstrahlung ($q_{br}$), synchrotron ($q_{syn}$) processes and inverse Comptonization
($q_{comp}$) due to soft synchrotron photons is given by
\begin{eqnarray}
q^-=Q^- \frac{c^{11}}{hG^4 M^3}=q_{br}+q_{syn}+q_{comp}
\label{qm}
\end{eqnarray}
where
\begin{eqnarray}
\nonumber
q_{br} &= &1.4 \times 10^{-27} \ n_e\,n_i T_e^{1/2}\,(1+4.4\times 10^{-10} T_e)\,
\,\,{\rm erg/cm^3/sec},\\
\nonumber
q_{syn} & = &\frac{2 \pi}{3 c^2} kT_e \, \frac{\nu_a^{3}}{R}\,\,\,{\rm erg/cm^3/sec},
\,\,\,q_{comp} = {\cal F}\,q_{syn}\,\,\,{\rm erg/cm^3/sec},
\,\,\,R=x\,GM/c^2,\\
\nonumber
{\cal F}& = &\eta_{1}
\left(1- \left(\frac{x_{a}}{3 \theta _{e}}\right)^{\eta _{2}} \right),\,\,\,
\eta _{1} = \frac{p(A-1)}{(1-pA)},\,\,\, p= 1 - \exp(-\tau_{es}),\\
A &=& 1 + 4 \theta_{e} + 16 \theta^{2}_{e},\,\,\,\theta_{e}=
kT_{e}/m_{e} c^{2},\,\,\
\eta _{2} =  1 - \frac{ln(p)}{ln(A)},\,\,\,x_{a}= h \nu_{a}/m_{e} c^{2},
\label{qvari}
\end{eqnarray}
when $\tau_{es}$ is the scattering optical depth, $\nu_a$ is the
synchrotron self-absorption cut off frequency.
Now following previous work\cite{rm10}, we solve the set of disk conservation equations
to obtain solution. We define a quantity called cooling factor, $f$, such that
\begin{eqnarray}
f=\frac{Q_{ie}-Q^-}{Q_{ie}},
\end{eqnarray}
which determines the efficiency of cooling in the flow.

\section{Results}\label{sol}

\begin{figure}[ht]
\includegraphics[height=.3\textheight,width=0.7\textwidth]{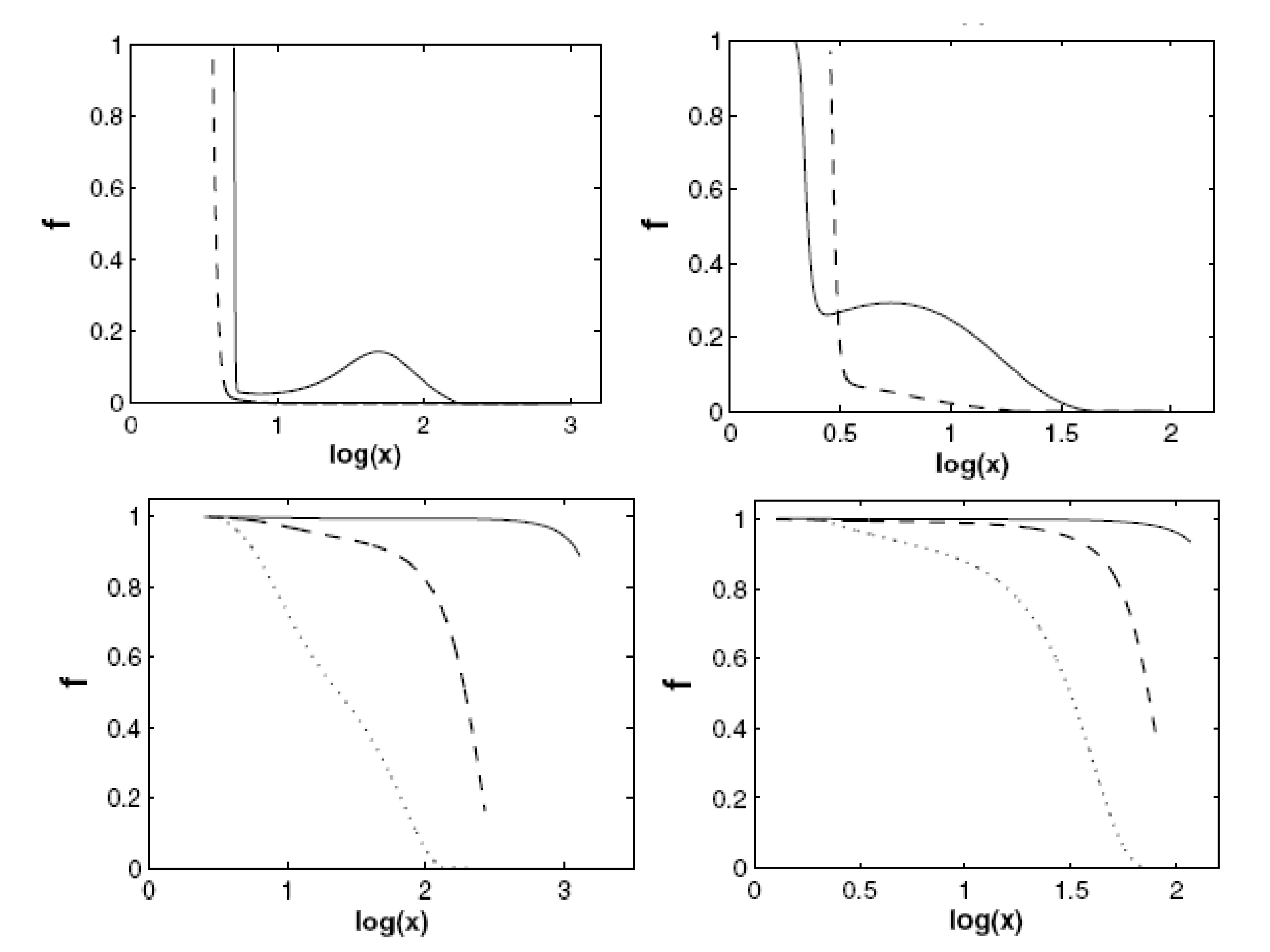}
\vskip-0.5cm
\hskip9.5cm
\caption{Top-Left: stellar mass ($M=10$), super-Eddington, $a=0$; Top-Right:
stellar mass ($M=10$), super-Eddington, $a=0.998$; Bottom-Left: super-massive
($M=10^7$), sub-Eddington, $a=0$; Bottom-Right: super-massive ($M=10^7$), 
sub-Eddington, $a=0.998$.
Solid, dashed curves in upper panels are for $\dot{m}=10,100$ Eddington rates
and Solid, dashed, dotted curves in lower panels are for $\dot{m}=0.01,0.1,1$
Eddington rates. For $a=0$, $\lambda=3.2$ and for $a=0.998$, $\lambda=1.7$.
\label{fig1}}
\end{figure}

We concentrate on two extreme cases: stellar mass black hole with super-Eddington 
accretion (StBSupA) and super-massive black hole with sub-Eddington accretion
(SuBSubA). While the 
former describes highly luminous X-ray sources (e.g. SS433), the later is for
low luminous AGNs and quasars (e.g. Sgr~A$^*$). For StBSupA, density of the
flow is higher than that of SuBSubA, which results in efficient cooling processes
therein compared to the later case. As a result the flow is cooler in StBSupA
than that in SuBSubA. Hence the difference in temperature between
protons and electrons in StBSupA ($\lsim 10$K) is smaller in the former case
compared to the later case ($\gsim 100$K). Figure \ref{fig1} shows that $f$
is very small in StBSupA until very close to the black hole, while in SuBSubA
it is very high in most of the inner disk region. However, in either of the cases,
flow appears hotter around rotating black holes compared to nonrotating ones.
This is because the specific angular momentum of the flow is smaller around
rotating black holes compared to nonrotating ones which results in a faster
infall and hence low residence time of the flow in the former case which does not allow
the cooling processes to complete before the flow impinges into the black hole.

It has already been understood that the under-luminous source Sgr~A$^*$ of mass $M=4.5\times 10^6$
accretes in a sub-Eddington accretion rate giving rise to a very low luminosity $L\sim 10^{33}$erg/sec.
Based on our model with $\dot{m}=10^{-5}$, $0.05\lsim\alpha\lsim 0.2$,
$4.9\times 10^{32}\lsim L\lsim 2.5\times 10^{33}$ only if $0.2\lsim a\lsim 0.5$.
This argues the black hole to be of intermediate spin.

\section{Conclusions}\label{dis}

We have the following punchline out of our two temperature, optically thin,
sub-Keplerian accretion disk.\\
$\bullet$ During infall, the flow governs much lower electron temperature ($\sim 10^{8-9.5}$K)
compared to proton temperature ($\sim 10^{10.2-11.8}$K), in the range of accretion
rate $10^{-2}\lsim\dot{m}\lsim 100$. This could explain hard X-rays and $\gamma$-rays
from AGNs and X-ray binaries.\\
$\bullet$ Weakly viscous flows are cooling dominated compared to their highly viscous
counter part of radiatively inefficient flows.\\
$\bullet$ The model flows transit from radiatively inefficient phase to
cooling dominated phase and vice versa, depending on the system, during infall.\\
$\bullet$ The model is able to reproduce a wide range of luminosities observed 
from under-fed AGNs and quasars (e.g. Sgr~$A^*$) to highly-luminous
X-ray sources (e.g. SS433), as well as highly-luminous quasars (e.g. PKS~0743-67).\\
$\bullet$ Based on our results Sgr~$A^*$ appears to be an intermediate spinning black hole
with the possible range of spin: $0.2\lsim a\lsim 0.5$. \\

\section*{Acknowledgments}
This work is partly supported by a project, Grant No. SR/S2HEP12/2007, funded
by DST, India.






\end{document}